%% file: main.tex
\documentclass[mathpazo]{csam}

\begin{document}
%%%%% title : short title may not be used but TITLE is required.
% \title{TITLE}
% \title[short title]{TITLE}
\title{Solving the $k$-sparse Eigenvalue Problem with Reinforcement Learning}

%%%%% author(s) :
% single author:
% \author[name in running head]{AUTHOR\corrauth}
% [name in running head] is NOT OPTIONAL, it is a MUST.
% Use \corrauth to indicate the corresponding author.
% Use \email to provide email address of author.
% \footnote and \thanks are not used in the heading section.
% Another acknowlegments/support of grants, state in Acknowledgments section
% \section*{Acknowledgments}
%\author[O.~Author]{Only Author\corrauth}
%\address{School of Mathematical Sciences, Beijing Normal University,
%Beijing 100875, P.R. China}
%\email{{\tt author@email} (O.~Author)}

% multiple authors:
% Note the use of \affil and \affilnum to link names and addresses.
% The author for correspondence is marked by \corrauth.
% use \emails to provide email addresses of authors
% e.g. below example has 3 authors, first author is also the corresponding
%      author, author 1 and 3 having the same address.

 \author[L. Zhou et~al.]{Li Zhou\affil{1},
       Lihao Yan\affil{2}, Mark A. Caprio\affil{2}, Weiguo Gao\affil{1}\comma\affil{3}, ~and Chao Yang\affil{4}\comma\corrauth}

 \address{\affilnum{1}\ School of Mathematical Sciences ,
          Fudan University,
          Shanghai 200433, P.R. China. \\
          \affilnum{2}\ Department of Physics ,
          University of Notre Dame,
          IN 46556, United States. \\
          \affilnum{3}\ School of Data Science ,
          Fudan University,
          Shanghai 200433, P.R. China. \\
          \affilnum{4}\ Computational Research Division,
          Lawrence Berkeley National Laboratory,
          CA 94720, United States.}
 \emails{{\tt lzhou11@fudan.edu.cn} (L.~Zhou), {\tt lyan4@nd.edu} (L.~Yan), {\tt mcaprio@nd.edu } (M.~Caprio), {\tt wggao@fudan.edu.cn} (W.~Gao),{\tt cyang@lbl.gov} (C.~Yang)}
% \footnote and \thanks are not used in the heading section.
% Another acknowlegments/support of grants, state in Acknowledgments section
%\section*{Acknowledgments}

%%%%% Begin Abstract %%%%%%%%%%%
\begin{abstract}
We examine the possibility of using a reinforcement learning (RL) algorithm to solve
large-scale eigenvalue problems in which the desired the eigenvector can be
approximated by a sparse vector with at most $k$ nonzero elements, where $k$ is relatively small compare to the dimension of the matrix to be partially diagonalized.
This type of problem arises in applications in which the desired eigenvector exhibits
localization properties and in large-scale eigenvalue computations in which the amount of
computational resource is limited.  When the positions of these nonzero elements can
be determined, we can obtain the $k$-sparse approximation to the original problem by
computing eigenvalues of a $k\times k$ submatrix extracted from $k$ rows and columns
of the original matrix.  We review a previously developed greedy algorithm for
incrementally probing the positions of the nonzero elements in a $k$-sparse approximate
eigenvector and show that the greedy algorithm can be improved by using an RL
method to refine the selection of $k$ rows and columns of the original matrix.
We describe how to represent states, actions, rewards and policies in an RL algorithm
designed to solve the $k$-sparse eigenvalue problem and demonstrate the effectiveness
of the RL algorithm on two examples originating from quantum many-body physics.
\begin{comment}
For the large sparse symmetric matrix eigenvalue problem, we are interested in approximating the solution by limiting the eigenvector be a k-sparse vector. For those matrices arising from physics problems, we can also observe eigenvector with a localization property. To overcome the  shortcoming of greedy method, we try to apply the reinforcement learning method to the k-spare eigenvalue problem. We define the states by all selected rows/columns and actions by a pair of indices $(p,q])$ where p is the index removing from state and q is the index adding to the state. The state and action spaces are tremendously large for defining $Q(s,a)$, in this case, the Q-table $Q(s,a)$ are approximated by linear combination of feature vectors. The reward for taking an action is defined by differences between the eigenvalues before and after taking the action. For the different feature vector, we develop two RL algorithm RL1 and RL2 for our problem, the numerical test on Li6Nmax and MBL20 matrix shows that the RL1 an RL2 algorithm achieve a better results than the baseline method and traditional sparse PCA method.
\end{comment}
\end{abstract}
%%%%% end %%%%%%%%%%%

%%%%% AMS/PACs/Keywords %%%%%%%%%%%
%\pac{}
\ams{xxxxx, xxxxx%The information of the AMS subject classification can be found in http://mathscinet.ams.org/msc/msc2010.html
}
\keywords{large-scale eigenvalue problem, quantum many-body problem, eigenvector localization, reinforcement learning, approximate Q-learning, stochastic sampling, high performance computing}

%%%% maketitle %%%%%
\maketitle

%%%% Start %%%%%%
\input{intro}
\input{greedy}

\input{rl}

\input{results}

\input{conclusion}
%%%% Acknowledgments %%%%%%%%
\section*{Acknowledgments}
This work was supported in part by the U.S. Department of Energy,
Office of Science, Office of Advanced Scientific Computing Research, Scientific
Discovery through Advanced Computing (SciDAC) program, and
the U.S.~Department of Energy, Office of Science, Office of Nuclear Physics,
under Award Number DE-FG02-95ER-40934. W. Gao is supported in part by National Science Foundation of China under Grant No. 11690013, 71991471,  U1811461.

\begin{comment}

\end{comment}

\bibliographystyle{abbrv}
\bibliography{ref}

%%%% Bibliography  %%%%%%%%%%

\end{document}

%% file: intro.tex
\section{Introduction}
\label{sec:intro}
Let $A$ be an $n \times n$ sparse symmetric matrix, where $n$ can be very large.
We are interested in solving the following problem
\begin{equation}
\min_{\|x\|_0 \leq k} \frac{x^T A x}{x^Tx},
\label{eq:evksparse}
\end{equation}
where $\|\cdot \|_0$ denotes the cardinality of a vector, i.e., the 
number of non-zero elements of a vector. A vector $x$
that satisfies $\|x\|_0 = k$ is called a $k$-sparse vector.
We will refer to \eqref{eq:evksparse} as a $k$-sparse eigenvalue
problem because the solution to \eqref{eq:evksparse} is 
the eigenvector associated with the algebraically smallest
eigenvalue of $A$ if the $k$-sparse constraint $\|x\|_0$ is
not imposed. When the $k$-sparse constrain is imposed, the solution 
to \eqref{eq:evksparse} can be obtained from the eigenvector of 
a submatrix of $A$ with at most $k$ rows and columns.

The $k$-sparse eigenvalue problem can also be more plainly stated as follows:
Select at most $k$ rows and columns of $A$ to form a submatrix $A_1$ such that
the algebraically smallest eigenvalue of $A_1$ is the smallest among 
all smallest eigenvalues of all submatrices of dimension at most $k$.
Note that we may replace the minimum in \eqref{eq:evksparse} by maximum if the eigenvalue of interest is the largest among all eigenvalues. 
This problem is related to the sparse principal component analysis (PCA) problem
in which $A$ is a covariant matrix of the form $A = B^TB$, and minimization
is replaced with maximization in \eqref{eq:evksparse}\cite{jolliffe2003modified,journee2010generalized,d2008optimal}.

If the eigenvector $\xopt$ associated with the algebraically smallest eigenvalue 
of $A$ has at most $k$ nonzero elements, it is the solution of \eqref{eq:evksparse}.
The positions of the nonzero elements of the eigenvector specify the rows
and columns of $A$ that defines $A_1$. 

If $\xopt$ has more than $k$ nonzero elements, it is not entirely clear how one can 
obtain the solution to  \eqref{eq:evksparse} efficiently or which rows and columns of $A$ should be extracted to form $A_1$ whose lowest eigenvalue yields the
minimum of the objective function in \eqref{eq:evksparse}. 
As we will show in Section~\ref{sec:results}, even if we can compute the smallest eigenvalue of $A$, simply taking $k$ rows and columns of 
$A$ corresponding to the $k$ largest components (in magnitude) of the corresponding eigenvector
does not necessarily yield the optimal solution to \eqref{eq:evksparse}.

The $k$-sparse eigenvalue problem is of particular interest when we try to solve a 
large-scale eigenvalue problem with a limited amount of computational resource.
One of the motivations originates from solving a quantum many body 
problem
\begin{equation}
\mathcal{H} \Psi = \Psi E,
\label{eq:schrodinger}
\end{equation}
where $\mathcal{H}$ is a many body Hamiltonian and $\Psi$ is an eigenfunction of $\mathcal{H}$ corresponding to the eigenvalue $E$. The lowest eigenvalue $E_0$ and its corresponding eigenfunction form the ground state of the many-body Hamiltonian~\cite{MQC,abinitioNuclearShellModel, nocoreshellmodel}. 

One way to solve \eqref{eq:schrodinger} is to expand $\Psi$ in terms
of a linear combination of a finite number of many-body basis functions 
known as Slater determinants in some well defined Hilbert space (often
referred to as a configuration interaction space), and
solve a projected linear eigenvalue problem in that subspace.
To obtain an accurate approximation to the solution of \eqref{eq:schrodinger},
the dimension of the configuration space may be prohibitively large.
Although significant progress has been made on solving this type of problem
on high performance computers using advanced sparse matrix techniques and
efficient iterative methods~\cite{SternbergSC08, topoeig, mfdneig, aktulga17,shao2018accelerating}, the size of the problems that we can 
currently solve is still limited by the amount of available computational resources.

However, it is well known that the ground state of many Hamiltonians have localization 
properties~\cite{Abanin,Ge2016}. This means that $\Psi$ can be well represented by a relatively small number of 
many-body basis functions. If these basis functions are known, we only need to diagonalize 
a Hamiltonian matrix of relatively small dimension.
However, in most cases, the optimal set of basis that allows us to give the best 
estimate of the ground state energy of the many-body Hamiltonian from a subspace (of many-body basis functions) of a fixed dimension (limited by computational resources) is not known. 
As a result, we need to learn how to identify the desired basis function and 
the corresponding rows and columns of $A$ to be extract as we try to solve the $k$-sparse 
eigenvalue problem.

We are interested in efficient methods for finding the solution of 
\eqref{eq:evksparse} without computing the eigenvalue and eigenvector
of $A$ directly, which can be prohibitively expensive.  We would like 
to identify the location of the $k$ non-zero components of the 
eigenvector associated with the smallest eigenvalue of $A$ if 
the eigenvector is indeed $k$ sparse (which we may not know 
a priori).  If we know where the locations of these 
component (but not the components themselves), we can then
diagonalize a $k$ by $k$ matrix $A_1$ to obtain the solution to \eqref{eq:evksparse}. In the case the eigenvector associated with the smallest
eigenvalue of $A$ is not strictly $k$-sparse, we would like to 
identify $k$ rows and columns of $A$ that yield a submatrix $A_1$, whose 
eigenvector associated with the smallest eigenvalue minimizes the objective function in \eqref{eq:evksparse} after it is padded with zeros.

In quantum chemistry and physics literature, several selected CI method have been developed to solve \eqref{eq:schrodinger} by selecting important Slater determinant basis~\cite{selectedCI,adaptiveSCI,wang2019coordinate}. These methods have been shown be competitive or sometimes better than Monte-Carlo based methods for sampling rows and columns of $A$ stochastically~\cite{alavi09,alavi13,LuWang20}.

In~\cite{greedy}, we developed a greedy algorithm to incrementally select rows and columns of $A$ to form $A_1$ based on either the residual or a component perturbation based selection criterion. We will review the basic idea of this approach in section~\ref{sec:greedy}. However, this greedy algorithm is far from optimal because the rows and columns of $A$ selected in the early stages of the greedy algorithm may not be the best ones for constructing $A_1$.
As we learn more about the matrix $A$ through the selection of additional rows and columns, it may become more clear that some of the rows and columns can be replaced by others that can contribute to lowering the smallest eigenvalue of $A_1$.

In this paper, we explore the possibility of using a reinforcement learning (RL) algorithm~\cite{silver2016mastering,sutton2018reinforcement} to improve the sampling of rows and columns of $A$ so that better approximation to the solution of \eqref{eq:evksparse} can be obtained. The RL algorithm uses a global expected reward function to guide the selection of rows and columns. The previously selected rows and columns may be removed to make room for new rows and columns that contribute more to lowering the smallest eigenvalue of $A_1$. This global reward
function is updated repeatedly to improve the selection policy. 

The basic elements of a RL algorithm is reviewed in section~\ref{sec:rl} where we also show how to formulate the search for an optimal set of rows and columns of $A$ to solve \eqref{eq:evksparse} as a RL process.  We provide some algorithmic details of the RL algorithm in the context of solving \eqref{eq:evksparse} also.
In section~\ref{sec:results}, we demonstrate the effectiveness of the RL approach with two numerical examples.

\section{Optimality Assessment}
Before we start to discuss methods for solving the $k$-sparse eigenvalue
problem, it is worth discussing first how to assess the optimality of an approximate solution.

The optimal solution of \eqref{eq:evksparse} is well defined,
if \eqref{eq:evksparse} is strictly $k$-sparse, meaning that the desired
eigenvector has at most $k$ nonzero elements.
We can reorder the elements of the eigenvector to have all nonzero elements appear in the leading $k$
rows, i.e.,
\[
Px =
\begin{bmatrix}
x_1 \\
0
\end{bmatrix},
\]
where $x_1 \in \mathbb{R}^{k}$ and $P$ the permutation matrix associated with such a reordering.
Consequently, we can reorder the rows and columns of the matrix $A$ so that
\begin{equation}
(PAP^T) (Px) =
\begin{bmatrix}
A_1 & A_{21}^T \\
A_{21} & A_2
\end{bmatrix}
\begin{bmatrix}
x_1 \\
0
\end{bmatrix}
=
\lambda
\begin{bmatrix}
x_1 \\
0
\end{bmatrix}
\label{eq:PAP}
\end{equation}
holds.

To obtain $x_1$, we just need to solve the eigenvalue problem
\begin{equation}
A_1 x_1 = \lambda x_1.
\label{eq:A1eig}
\end{equation}
Assuming $\lambda$ is simple, we can verify the optimality of the solution by comparing the solution to \eqref{eq:A1eig} to that of 
\eqref{eq:PAP} (which may be costly to obtain.) 

However, when the desired eigenvector is not strictly $k$-sparse, the matrix $A_1$ 
that yields the best approximate eigenvalue
may not be the submatrix obtained by extracting rows and columns of $A$ associated with $k$ largest  (in magnitude) elements of the desired eigenvector of $A$. 

In theory, one can obtain the optimal solution of \eqref{eq:evksparse} by
enumerating all possible combinations of $k$ rows and columns out of $n$ rows and
columns of $A$, solving each $k$-dimensional eigenvalue problem
and choose the optimal among solutions to all $k$-dimensional eigenvalue problems.
However, in practice, this approach is prohibitively expensive even for problems 
with a moderate $n$ and $k$ because $\left( \begin{array}{c} n \\ k \end{array} \right)$
can be extremely large.

Let us denote the smallest eigenvalue of the $A_1$ matrix obtained by extracting rows and columns of $A$ associated with $k$ largest (in magnitude) elements of the desired eigenvector of $A$ by
$\lambda^b$.

We say that an approximate solution to \eqref{eq:evksparse} is a "good" solution if the smallest eigenvalue $\theta$ of the $A_1$ matrix selected from the RL learning algorithm presented in this paper is less than or equal to $\lambda^b$.

The true optimal solution lies between $\lambda^b$ and $\lambda$. Clearly, we would like to obtain an approximate solution that is as close as possible to $\lambda$. However, how close $\theta$ is to $\lambda$ depends on $k$  and how localized the eigenvector $x^*$ is.

%% file: greedy.tex
\section{Greedy Probing}
\label{sec:greedy}
One way to solve the $k$-sparse eigenvalue problem is to use the greedy algorithm presented in~\cite{greedy}.
The basic algorithm can be summarized as follows
\begin{enumerate}
\item We select a subset of the indices $1,2,...,n$ denoted by
$\mathcal{S}$ that corresponds to ``important'' rows and columns of $A$.
In the configuration interaction method for solving quantum many-body eigenvalue
problems, this subset may correspond to a set of many-body basis functions
produced from some type of basis truncation scheme~\cite{selectedCI,heatBathCI,adaptiveSCI}.
\item Let $A_1$ be a submatrix of $A$ that consists of rows and columns 
defined by $\mathcal{S}$. Assuming the size of $\mathcal{S}$ is
small relative to $k$ (and $n$), we can easily compute the desired eigenpairs $(\lambda_1, x_1)$ of $A_1$, i.e., $A_1 x_1 = \lambda_1 x_1$.
\item We take $\lambda_1$ to be the approximation to the smallest eigenvalue of $A$.
The approximation to the eigenvector of $A$ is constructed as
$\hat{x} = P^T \begin{bmatrix} x_1^T & 0 \end{bmatrix}^T$. To assess the accuracy of the computed eigenpair $(\lambda_1,\hat{x})$,
we compute the full residual $r = A \hat{x} - \lambda_1 \hat{x}$.
\item If the norm of $r$ is sufficiently small, we terminate the computation
and return $(\lambda_1,\hat{x})$ as the approximate solution. Otherwise,
we select some additional rows and columns of $A$ using an appropriate selection criterion to augment $A_1$ and repeat steps 2--4 until the dimension of $A_1$ is $k$.
\end{enumerate}

Note that steps 2--4 of the above algorithm do not need to be repeated if $k$ rows and and columns of $A$ are selected all at once in step 4. But for large problems in which $k$ can still be relatively large, it is generally computationally more efficient to select a few rows
and columns at a time to construct $A_1$ incrementally. Such a scheme often yields better approximation also because it may not be clear in advance which $k$ rows and columns of $A$ we should select.

We now discuss a few strategies for selecting rows and columns of $A$ incrementally to obtain an $k\times k$ submatrix $A_1$ that yields a good $k$-sparse approximation to the desired eigenpair of $A$.

\subsection{Residual based approach}
Without loss of generality, we take $\mathcal{S}$ to be the leading $n_1 \ll n$
rows and columns of $A$ so that we can partition $A$ as
\begin{equation}
A =
\begin{bmatrix}
A_1 & A_{21}^T \\
A_{21} & A_2 
\end{bmatrix}.
\label{eq:Apart}
\end{equation}
We now discuss how to select additional ``important'' rows and columns outside of the subset $\mathcal{S}$ to obtain a more accurate approximation of the
desired eigenvector of $A$.

Suppose $(\lambda_1, x_1)$ is the computed eigenpair of the submatrix $A_1$
that serve as an approximation to the desired eigenpair $(\lambda,x)$.
By padding $x_1$ with zeros to form
\begin{equation}
\hat{x} =
\begin{bmatrix}
x_1 \\
0
\end{bmatrix},
\label{eq:xhat}
\end{equation}
we can assess the accuracy of the approximate eigenvector $\hat{x}$
in the full space by computing its residual
\begin{equation}
r = A \hat{x} - \lambda_1 \hat{x}
=
\begin{bmatrix}
0 \\
A_{21} x_1
\end{bmatrix}
\equiv
\begin{bmatrix}
0 \\
r'
\end{bmatrix}.
\label{eq:residual}
\end{equation}

The first greedy scheme for improving the accuracy of $x_1$ is to select
some row indices in $\{1,2,...,n\} \setminus \mathcal{S}$ that
correspond to components of $r'=A_{21}x_1$ with the largest magnitude.
These indices, along with $\mathcal{S}$, yield an augmented $A_1$
from which a more accurate approximation to $(\lambda, x)$ can be obtained.

\subsection{Perturbation based approach}
Another greedy probing algorithm can be developed by a component-wise perturbation analysis in which one component in the zero block of  $\hat{x}$ defined by \eqref{eq:xhat} is perturbed to make 
\begin{equation}
\tilde{x} = \left(
\begin{array}{c}
x_1 \\
\gamma_j e_j
\end{array}
\right)
\label{eq:xpert1}
\end{equation}
a better approximation to the desired eigenvector. The 
analysis presented in~\cite{greedy,PToriginalpaper} shows that $\gamma_j$ can be estimated to be 
\begin{equation}
\gamma_j \approx \frac{e_j^T A_{21} x_1}{\lambda_1 - e_j^TA_{2} e_j}.
\label{eq:gamma}
\end{equation}
We choose $j \in \{1,2,...,n\} \setminus \mathcal{S}$ that yields large values of $|\gamma_j|$'s to argument $\mathcal{S}$ incrementally in the greedy algorithm to 
ultimate obtain a
$k\times k$ matrix $A_1$ from which an approximation the desired eigenpair can be computed.

\subsection{Fast Update}
Both the residual 
and component-wise perturbation analysis based greedy probing algorithm require computing 
$A_{21} x_1$. When 
the dimension of $A$ is large, which is the case we are primarily interested in, this computation can be prohibitively costly, especially if we have to perform it each time $A_1$ is augmented~\cite{heatBathCI}. The cost of the computation can be reduced if we exploit the sparsity structure of $A_{21}$, i.e., we only multiply nonzero rows of $A_{21}$ with $x_1$. The identification of these nonzero rows is problem dependent. For quantum many-body problems arising from chemistry, several strategies have been developed to perform this update efficiently~\cite{OLSEN1990,heatBathCI,semiHeatBathCI}.

To avoid recomputing $A_{21} x_1$ whenever 
$A_1$ is augmented in multiple stages, we use a simple updating scheme described below.

We denote the partitioned blocks of $A$ in the $m$th stage of the greedy algorithm by $\at{m}$ and $\as{m}$  respectively.

Let us partition the matrix $\at{m+1}$ by 
\begin{equation}
\at{m+1} = \left( 
\begin{array}{cc}
\at{m} & B^T \\
B & C
\end{array}
\right),
\label{eq:A1a}
\end{equation}
where the $B$ and $C$ blocks correspond to 
newly added rows and columns
in the $m+1$st stage.
The eigenvector of $\at{m+1}$, denoted by $\xw{m+1}$ can
be partitioned conformally with that of $\at{m+1}$, i.e.,
\begin{equation}
\xw{m+1} 
= \left(
\begin{array}{c}
\xwh{m+1} \\
y
\end{array}
\right).
\end{equation}

Let $\as{m+1}$ be the $(2,1)$ block of $A$ after $\at{m}$ is augmented to $\at{m+1}$. We can partition this matrix conformally with the way $\xw{m+1}$ is partitioned, i.e.,
\begin{equation}
  \as{m+1}
 = \left(
 \begin{array}{cc}
   \ahs{m}  &  E 
 \end{array}
 \right),
\end{equation}
where $\ahs{m}$ is a submatrix of $\as{m}$ after the submatrix $B$ block is removed.
As a result, the vector
$\as{m+1}\xw{m+1}$, which is required in the $m+1$st stage of both the residual and perturbation based greedy probing algorithms for selecting additional rows and columns, can be computed as
\begin{equation}
\as{m+1} \xw{m+1} = 
\ahs{m} \xwh{m+1} + 
E y.
\label{eq:a21upd}
\end{equation}
Note that $\ahs{m}$ contains a subset of rows of $\as{m}$ in \eqref{eq:Apart} before $\at{m}$ is augmented. Therefore, if $\xwh{m+1}$ is close to $\xw{m}$, we may use components of $\as{m}\xw{m}$, which have already been computed, in place of those in  
$\as{m+1} \xw{m+1}$ in 
\eqref{eq:a21upd}. The only additional computation we need to perform is $Ey$.
Because $\|\xwh{m+1}\| < 1$ when $y\neq 0$, whereas $\|\xw{m}\|=1$, to account for the difference in scale, we multiply selected components of the previously computed $\as{m} \xw{m}$ by a scaling factor $\nu = \|\xwh{m+1}\|$ 
before it is combined with components of  $Ey$ to yield an estimation for components of $\as{m+1} \xw{m+1}$.

%% file: rl.tex
\section{Reinforcement Learning}
\label{sec:rl}
One of the main issues with the greedy algorithm reviewed in the previous section
is that the algorithm terminates when the dimension of $A_1$ reaches $k$, but the selected rows and columns may be far from optimal at that point.

In this section, we examine how to use a RL algorithm to improve the incremental greedy probing method.
In particular, we will examine a procedure that performs the greedy selection
repeatedly after the dimension of $A_1$ reaches $k$. This requires
us to remove some rows and columns from $A_1$ in order to bring in new rows
and columns of $A$ that can yield a smaller $\lambda_1$.
We will also discuss criteria and strategies for selecting rows and columns.

Just like the greedy probing scheme, the RL algorithm we examine only needs to access a small fraction 
of $A$ in every step. Therefore, we do not need to store the entire $A$ in advance.

\subsection{Reinforcement Learning Basics}
In a reinforcement learning algorithm, an agent is trained to take a sequence
of actions from a state to reach other states with the ultimate goal
of achieving a predefined objective.

In our case, the states are simply different combinations of $k$ rows and
columns selected from $A$ to form $A_1$. 
The state that solves the k-sparse eigenvalue problem
is called the optimal state that the agent wants to reach after taking a sequence of actions from a starting state that is non-optimal. Each action corresponds
to removing one or more rows and columns from $A_1$ and selecting some additional
rows or columns from $A$ to replace the removed rows and columns.
The effectiveness of each action is measured by a reward function
which takes the state and action pair as the input and gives a score 
for such a pair, which can, for example, be the change in the smallest 
eigenvalue of $A_1$ or some other metric. 

Because the reward associated with each state/action pair is local,
taking an action that gives the largest reward at a particular 
state does not necessarily lead to an optimal strategy globally.
For example, even when replacing some rows and columns of $A_1$ 
with others from $A$ can lead to a smaller decrease or even an 
increase in the the smallest eigenvalue of $A_1$, the selected 
rows and columns may be important in minimizing the smallest eigenvalue 
of $A_1$ when combined with other rows and columns selected in
subsequent steps.

We decide which action to take by following a policy prescribed by a function $Q(s,a)$, which can be viewed as the sum of discounted future rewards the agent can expect on average after it reaches the states $s$ and chooses to take the action $a$.
This $Q(s,a)$ function is constructed and refined dynamically in a multi-episode  training process. The policy defined in terms of $Q(s,a)$ is related to, but not completely determined by the local reward.  During a training process, we may not want the agent to always take the action that yields the largest reward at that moment. Instead,  we may want the agent to have the opportunity to explore different possibilities to improve $Q(s,a)$, which can be viewed as giving a more global assessment of the $(s,a)$ pair towards the goal of solving \eqref{eq:evksparse}. 

\begin{comment}
Therefore, it is useful to record the reward associated with
state/action pairs in some way so that a more globally optimal
strategy can be developed by examining the global reward of
the actions from a particular state. 
For example, one way of approximating the global optimal reward 
may be using some heuristics. \LY{What does heuristics mean?}
This globally optimal reward of each state/action pair is often called the 
q-value in RL literature.
The reinforcement learning scheme that learns the qvalues of
each state/action pair directly is called qlearning. 

We may then define an optimal policy that uses the 
qvalues for each state to select the optimal actions to take. 
Guided by the policy, the agent can perform a sequence of actions
and reach the optimal state that solves our goal. 
\end{comment}

\begin{comment}
% above is a complete loop: now goes deeper into the loop
In our problem, the matrix \(H\) we are dealing with
is prohibitively large, so the agent does not know 
the rewards of particular actions before the action is taken. Once a action
is taken to reach a new state, the agent learns the local reward of that action.
The local reward is then be used to train the qtable or qfunction.
Through trial and error, the agent learns. 
\LY{is this explanation good enough?}
\end{comment}

A reinforcement learning algorithm that produces an optimal policy defined by an optimal $Q(s,a)$ is 
known as a $Q$-learning algorithm~\cite{Watkins:1989,watkins1992qconvergence,sutton2018reinforcement}.
The construction of $Q(s,a)$ can be exact or approximate.  An exact $Q(s,a)$ may be obtained and tabulated if  the number of state and action pairs is relatively small. When the number of state and action pairs is too large, we typically need to develop a way to approximate $Q(s,a)$. Such an approximation can be as simple as taking a linear combination of important factors or features of the $(s,a)$ pair~\cite{bucsoniu2011approximate,tsitsiklis1996feature,barreto2020fast}.
It can also be represented as by a deep neural network if a nonlinear parametrization can better capture the behavior of the function~\cite{silver2016mastering,lecun2015deep,mnih2015human} .

In this paper, we approximate $Q(s,a)$
by using {\em feature-based representation} of state/action pairs.
For a given $(s,a)$ pair, such a representation can be expressed in terms of a 
vector $f(s,a) \in \mathbb{R}^m$. The function $Q(s,a)$ can be  expressed in terms of a linear combination of feature components,
i.e.,
\begin{equation}
Q(s,a) = \sum_{i=1}^m w_i f_i(s,a),
\label{eq:qsa}
\end{equation}
where $w_i$ is a weight factor.
We will define $s$, $a$, $f(s,a)$, $Q(s,a)$ and how they are represented in greater details
in the next section.

\subsection{State, Action, Reward and Policy Representation}
\label{section:RLdefinitions}
To solve \eqref{eq:evksparse} by RL, we train an agent to 
iteratively refine the selection of rows and columns of $A$ 
to be included in the $k\times k$ submatrix $A_1$. The training procedure allows us to construct and update
an expected (global) reward
function $Q(s,a)$ associated with each state and action pair through the update of a set of feature weights.

After the training is completed, the optimal policy allows us to select rows and columns of $A$ in an optimal fashion to solve \eqref{eq:evksparse}. We will now discuss how state, action, reward and policy are defined specifically for a RL algorithm designed to solve the $k$-sparse eigenvalue problem \eqref{eq:evksparse}.

As we indicated earlier, each state corresponds to a set of $k$ 
rows and columns of $A$ used to form $A_1$. The order of these 
rows and columns is not important.
The environment for our agent is the set of all possible states.

There are multiple ways to represent a state. One possibility is to use a 
size-$n$ binary indicator vector 
    \[ s = (0,1,1,0,\dots) \]
    to mark which rows and columns are not included (indicated by 0's) and which are included (indicated by 1's). An alternative representation is simply a set of indices of the rows/columns of $A$ that are included in $A_1$, 
    e.g., 
    \[s = \{2,3,...\}.\]
The set of all possible states, which can be very large,
is denoted by $\mathbb{S}$.

For simplicity, let us first define an action as removing a row/column from $A_1$
and adding a new row/column from $A \setminus A_1$ to $A_1$. This is equivalent to
removing an index \(p\) from $s$ and adding a new index $q$ that is not in $s$.
We can denote such an action by a pair of indices, i.e., $a = (p,q)$. 
The space of all actions for a particular state, which can be very large, is denoted by $\mathbb{A}$.

Each action is associated with a local reward. One natural choice of the
reward function is 
    \begin{equation}
        r= \theta - \theta',
    \label{eq:reward}
    \end{equation}
where $\theta$ is the smallest eigenvalue of $A_1$, 
associated with
the state $s$ before the action is taken, and $\theta'$ is 
the smallest eigenvalue of $A_1$ associated with the state $s'$ after 
the action is taken.

The true impact of the action $a$ may not be known until some future actions are taken. The global expected reward of the state action pair $(s,a)$, which allows us to devise an effective policy towards finding the optimal 
$A_1$, is defined by a function, 
$Q(s,a):\mathbb{S} \times \mathbb{A} \mapsto \mathbb{R}$.
As we indicated in \eqref{eq:qsa}, in an approximate $Q$-learning scheme, $Q(s,a)$ can be defined in terms of a weighted average of feature components associated with a feature vector defined for
each \((s,a) \).
For each state action pair $(s,a)$, where the action $a$ involves taking $p$ out of $s$ and adding $q$ into the new $s'$, $f_i$ can be  defined as:
\begin{equation}
    f_i(s,a) = 
\delta_{i,q} - \delta_{i,p}, \ \ \mbox{for $p\in s$ and $q \notin s$},
\label{eq:fidiff}
\end{equation}
where $\delta_{i,j} = 1$ if $i = j$ and 0 otherwise.
Note that when $p\in s$ and $q \notin s$ hold, the vector 
\(f(s,a) \) contains only two nonzero elements.
As a result, the value of $Q(s,a)$ depends on the difference between $w_q$ and $w_p$ for a specific $(s,a)$ pair. 

Another way to define $f(s,a)$ is to set 
\begin{equation}
f_i(s,a) =
\left\{
\begin{array}{cc}
1 & \mbox{if $i\in s$ and $i\neq p$, or if $i \notin s$ and $i = q$.}\\
-1 & \mbox{if $i\in s$ and $i = p$} \\
0 & \mbox{otherwise}. 
\end{array}
\right.
\label{eq:fisum}
\end{equation}
In this case,  $Q(s,a)$ is simply the sum of weights associated with elements of $s'$ and $-w_p$ where $s'$ is the state achieved by $s$ taking the action $a=(p,q)$. 
It provides a measure of how important the rows and columns in $s'$ are collectively.

\begin{comment}
We use the optimal $Q(s,a)$ values to find the action that leads
to the optimal state regardless of what state we are in right now.
\end{comment}

Note that the weights in \eqref{eq:qsa}  are themselves independent of $s$ and $a$. They give, to some extent, a global ranking of rows and columns of $A$ in terms of their importance in being selected to form $A_1$. They are updated after an action $a$ has been taken.
As we will show below, the update of 
$w_i$'s makes use of the local reward
$r$. It balances the previous values and their changes through a learning rate $\alpha$.  It also take into account future rewards resulting from taking a subsequent action $a'$
from the new state $s'$. We can view the update of $w_i$'s as taking a gradient descent step towards minimizing the difference between the current $Q(s,a)$ and an optimal (but unknown) $Q(s,a)$.  We will examine this optimization point of view in section~\ref{sec:detail}.

Before the weights $w_i$'s are updated, we need to decide which action to take at each $s$. In the RL literature, a rule for making such a decision is called a \textit{policy}, which is defined to be a function \(\pi(s):\mathbb{S} \mapsto \mathbb{A} \)
~\cite{sutton2018reinforcement}.
Intuitively, the policy 
should be designed to select actions that can lead to the largest expected reward $Q(s,a)$, and ultimate reach an optimal state  $s^{\ast}$ that solves \eqref{eq:evksparse}, i.e., $\pi(s)$ is defined as

\begin{equation}
  \pi(s) = 
   \argmax_a Q(s,a), 
  \label{eq:policy}
\end{equation}

However, due the large number of possible actions the agent can take at each $s$, this optimization problem may not be  easy to solve. One way to overcome this difficulty is to obtain a nearly optimal solution by considering a subset of actions that are likely to produce a larger $Q(s,a)$ value. This choice also depends on how $Q(s,a)$ is defined. We will examine two specific policies in section~\ref{sec:policy}.

Because $Q(s,a)$ is a function 
that is learned over time during the training process, the policy also 
evolves during the training process.
Upon the completion of the training process, we obtain, in principle, an optimal \(Q^*\), which corresponds to a set of optimal weights $w_i$'s 
associated with all rows and columns of $A$. 
This optimal set of $w_i$'s provides a ranking of all rows and columns of $A$ in terms of their importance in contributing to the solution of \eqref{eq:evksparse}. We form $A_1$ by simply selecting rows and columns of $A$ associated with largest $w_i$'s.

\subsection{A Basic RL Algorithm for Solving a $k$-sparse Eigenvalue Problem}

Using notation established above, we outline the basic steps of the RL algorithm for solving the $k$-sparse eigenvalue problem \eqref{eq:evksparse}.
\begin{algorithm}[H]
	\begin{algorithmic}[1]
		\State {\bf Input}: matrix $A$ (efficiently represented), the sparsity level $k$;
		\State {\bf Output}: Approximate solution $(\theta, x)$ of \eqref{eq:evksparse}, where $x$ contains at most $k$ nonzero elements 
		\State Initialize the weight $w_i$ (thus $Q(s,a)$ for all $s$'s and $a$'s;
		\For {episode=1,2,...,epmax} 
			\State Select an initial state $s$;
			\While{not done} 
				\State \begin{minipage}[t]{5.2in}Choose and take a action according to the policy $\pi(s)$ based on $Q(s,a)$ values for all possible $a$'s;
				\end{minipage}
				\State Evaluate the local reward $r$;
				\State \begin{minipage}[t]{5.2in}Use $r$, a learning rate $\alpha$, a reward discount rate $\gamma$ to update $Q(s,a)$ implicitly by updating $w$;\end{minipage}
			\EndWhile
		\EndFor
		\State Select the best state using the training results;
	\end{algorithmic}
	\caption{The Basic RL algorithm}
	\label{alg:basicRL}
\end{algorithm}

The algorithm consists of two nested loops. The outer loop runs over a number of training episodes.  Each episode consists of a number of inner loop iterates in which an agent transitions from one state to another by following a search policy to take a sequence of actions. Each action is followed by a calculation of a local reward and the update of $Q(s,a)$ (through the update of feature weights).  We will provide details of state and weight initialization, search policy, reward calculation, $Q(s,a)$ update and termination criteria in the next section.

\subsection{Algorithmic Details}
\label{sec:detail}
In this section, we give some algorithmic details of Algorithm~\ref{alg:basicRL} that can lead to an efficient implement RL method for solving the $k$-sparse eigenvalue problem.

\subsubsection{State and Weight Initialization}
The first episode of RL begins when the greedy algorithm described in ~\ref{sec:greedy} produces
a $k \times k$ $A_1$ matrix.  The set of row and column indices of $A$ that have been selected to form $A_1$ defines the initial state $s$.

We initialize the weight $w_i$'s as follows.  If $i\in s$, we set $w_i = \beta_1 |e_i^T x_1|$, where $x_1$ is the eigenvector associated with the desired eigenvalue of $A_1$, and $\beta_1$ is a normalization factor.
If $i\notin s$, we set $w_i = \beta_2 |\gamma_i|$, where $\beta_2$ is another normalization factor, if $\gamma_i$ has been computed from \eqref{eq:gamma} as the result of perturbation analysis performed in the greedy algorithm. Otherwise, $w_i$ is simply set to 0. The normalization factors $\beta_1$ and $\beta_2$ are chosen to ensure $e_i^T x_1$ and $\gamma_i$ are on the same scale. They depend on the dimension of the original problem ($n$), and the desired sparsity of the eigenvector $k$.

Upon the completion of each RL episode, we need to reinitialize the state $s$. The weights are not reinitialized because they represent the ``global'' ranking of the rows
and columns of $A$ that are learned over training episodes.

The state reinitialization strategies are not unique. In this paper, we present two state reinitialization 
methods used in two variants of the RL algorithm.
In the first method, we initialize $s$ by first selecting the rows that have the largest weights and
then replace the rows in \(s\) that have the lowest weights with the
rows outside of \(s\) that have the largest weights. In the second method, we simply initialize $s$ to be 
the last state reached in the previous episode.

\subsubsection{Search Policy}
\label{sec:policy}
We now discuss how to decide which action to take when we reach the state $s$ during a training procedure. This decision defines a policy function $\pi(s)$. As we indicated earlier, an intuitive policy is to take the action that maximize the expected (global) reward, i.e., to take the action that maximizes $Q(s,a)$, as described by \eqref{eq:policy}.
The implementation of such
a policy depends on how $Q(s,a)$ is defined in terms of $w_i$'s, which in turn depends on how feature vectors are defined for each $(s,a)$ pair. 

It may appear that solving \eqref{eq:policy} is difficult due to the large number of potential choices of actions. By limiting our search range to states that correspond to largest values of $w_i$'s determined in previous training episodes, solving \eqref{eq:policy} becomes tractable.

When $f(s,a)$ is defined by \eqref{eq:fidiff}, $Q(s,a)$ is determined by $w_q - w_q$, where $p$ is the row/column to be removed and $q$ is the row/column to be added in $a$. Therefore, to solve the optimization problem \eqref{eq:policy} so that the best action can be determined, we just need to examine all pairwise difference $w_q - w_p$, for $p\in s$ and $q\notin s$, and choose the action that yields that largest difference. If $w_p$'s
and $w_q$'s are sorted in advance, we can simply take the $p$ that yields the smallest $w_p$, and $q$ that corresponds to the largest $w_q$.

An alternative search policy is to take into account both $w_i$ and some additional information in the determination of the best action to take from the state $s$.
We use the value of $w_i$'s for $i\in s$ to select a set of rows in $s$ to be potentially removed.
We denote this set of indices by $S_1$.
Because the values of $w_j$'s for $j \notin s$ may be far from optimal, we use an alternative metric to select potential rows and columns outside of $s$ to be added to $s'$.

There are a number of alternative metrics we can use. For example, we can use the magnitude of 
\begin{equation}
c_j = e_j^T A_{21} x_1,
\label{eq:cj1}
\end{equation}
where $x_1$ is the eigenvector associated with the smallest eigenvalue of $A_1$ defined by $s$. Instead of examining all rows of $A_{21}x_1$, which can be costly, we can examine a subset of these elements by sampling a few rows of $A_{21}$. The sampled rows are determined by the nonzero structure of $A_{21}$ and the magitude of the nonzero matrix elements.  We select a subset of rows
that correspond to elements of $A_{21}x_1$ with the largest magnitude, and place the global indices of these rows in a set $S_2$.

Instead of choosing $p = \argmin_{i\in S_1} w_i$ to be removed from $s$ and replacing it with $q = \argmax_{j \in S_2} c_j$, we go through all pairs of $(i,j)$ for $i\in S_1$ and $j \in S_2$, and compute the smallest eigenvalue $\theta'$ of 
the updated matrix $A_1$ obtained by replacing row $i$ with row $j$. 
We choose the action $a = (p,q)$ as 
the first pair that satisfies
\begin{equation}
\theta' < \theta\cdot (1-\tau \cdot \epsilon),
\label{eq:explore}
\end{equation}
where $\tau$ is an exploration rate and $\epsilon$ is a uniformly distributed random number in the interval $(0,1)$ ($U(0,1)$). Such a probablistic selection strategy makes it possible for the agent to escape from 
a locally optimal policy and reach a better state.  Effectively, the set of actions defined by 
$(p,q)$'s for $p\in S_1$ and $q \in S_2$ form an \textit{active space} from which a good action can be 
determined via exploitation and exploration. We refer to a search policy that selects an action from
an active space as an \textit{active space} based search policy. This approach is similar to the complete active space self-consistent field method used in quantum chemistry~\cite{ROOS1980}. We summarize such a policy in 
Algorithm~\ref{alg:policy}.

In addition to using \eqref{eq:cj1} as a criterion for selecting candidate rows/columns to be added to $s'$, we may also use
\begin{equation}
    c_j = \frac{|e_j^T A_{21}x_1|}{|\theta - e_j^T A_2 e_j|},
    \label{eq:cj2}
\end{equation}
which results from an estimation of the magnitude of the $j$th eigenvector component derived from first order perturbation analysis. 

Yet, another way to assess
how important row/column $j \notin s$ is in achieving the goal of solving \eqref{eq:evksparse} is to estimate the amount of change in the desired eigenvalue we can realize by bringing $j$ into $s$. The estimation can be obtained by computing the smallest eigenvalue of the $2 \times 2$ matrix
\begin{equation}
A'=
\begin{bmatrix}
\theta & x_1^T A_{12}e_j \\
e_j^TA_{21}x_1 & e_j^TA_2e_j\\
\end{bmatrix},
\label{eq:Ap}
\end{equation}
and subtracting it from $\theta$.
\begin{algorithm}[H]
	\begin{algorithmic}[1]
		\State {\bf Input}: the state $s$, $Q(s,a)$ (represented by weights $w_i$'s), $c_i$'s determined by \eqref{eq:cj1} or \eqref{eq:cj2}, exploration rate $\tau$; 
		\State {\bf Output}: the action pair $(p,q)$

		\State Construct $S_1$ to include $i\in s$ that have the smallest $w_i$'s;
		\State Construct $S_2$ to include $j\notin s$ that have the largest $c_j$'s;
		\For{ $j=1, 2, ..., |S_2|$}
			\For {$i=1, 2, ..., |S_1|$}
				\State $s'=((s \setminus S_1[i]) \cup S_2[j])$  
				\State Compute the smallest eigenvalue $\theta'$ of $A(i\in s', i\in s')$;
				\State Generate a random number $\epsilon \sim U(0,1)$;
				\If  {$\theta'$ \textless $\theta \cdot (1-\tau\cdot\epsilon)$}
					\State Let $p=S_1[i]$, $q=S_2[j]$ and output the action $(p,q)$.
					\State Caluate the reward and update $w$. 
				\EndIf
			\EndFor
		\EndFor		
	\end{algorithmic}
	\caption{Active space based search policy}
	\label{alg:policy}
\end{algorithm}

To improve efficiency, it is not necessary to generate new $S_1$ and $S_2$ for each state. Whenever an action $(p,q)$ is taken, we can update $S_1$ by  deleting and adding $q$. We update $S_2$ by deleting $q$ and marking all elements of $S_2$ 
that have already be considered be in previous action section steps.

\subsubsection{Local reward}
\label{sec:reward}

Local reward measures the immediate progress the agent makes towards the final goal of solving \eqref{eq:evksparse} after taking an action $a$ defined by \eqref{eq:reward}. One natural way to define such a reward is to calculate the difference between desired eigenvalues computed before and after the action is taken. If $\theta'$ is the eigenvalue of the update $A_1$ matrix obtained by taking an action $a$, the local reward can be calculated as $\theta-\theta'$. 
Local rewards are used to update the $Q(s,a)$ which measures the global expected reward 
associated with a state-action pair. When $Q(s,a)$ is defined as a weighted sum of components of a feature vector, updating $Q(s,a)$ is equivalent to updating the weighting factors $w_i$'s. To update several weight factors simultaneously, it is convenient to partition the local reward among rows and columns that constitute the state $s$. Let $x_1 = (\xi_1, \xi_2, ..., \xi_k)^T$ be the eigenvector of $A_1$ associated with the eigenvalue $\theta$. We  can then partition $r$ as 
\begin{equation}
r = \sum_i r_i,
\label{eq:rsum}
\end{equation}
where $r_i = r \xi_i^2$.

\subsubsection{Update $Q(s,a)$}
% updating rule
Once the agent takes an action $a$ from $s$ to reach $s'$, we update the function $Q(s,a)$ by modifying the weights $w_i$'s using the following
scheme:
\begin{equation}
    % [TODO: SUBJECT TO ADJUSTMENTS]
        w_i^{m+1} \leftarrow w_i^m + \alpha \cdot \left[ r + \gamma \cdot \max_{a'} Q(s',a') - Q(s,a)\right] \cdot f_i(s,a)
    \label{eq:updateRule}
\end{equation}
where \(m\) is the (inner) iteration index, \(\alpha \) is known as the learning rate,
\(r\) is the local reward resulting from taking the action $a$, e.g., defined by \eqref{eq:reward},
\(\gamma\) is known as a  discount rate for a reward to be collected from a future action. Recall that $f_i(s,a)$ is the $i$th component of a feature vector associated with state action pair $s$ and $a$. 
The term \(\max_{a'}Q(s',a')\) is the maximum future rewards that can be collected after action \(a\) is taken. 
%It can again be calculated as the maximum difference between $w_q$  and $w_p$ for $p \in s'$ and $q\notin s'$.
% TODO: minimizing error, and show the rationale for the difference term and why we has a time

This updating formula can be viewed as a gradient descent step in the optimization of $Q(s,a)$~\cite{bucsoniu2011approximate}.
Suppose the optimal value of a particular $(s,a)$ pair is \(Q^*(s,a)\), which we do not know in advance. The goal of a RL training process is to minimize the distance between the existing $Q(s,a)$ value and $Q^\ast(s,a)$ with respect to $w_i$'s. 

Since we represent $Q(s,a)$ by
\eqref{eq:qsa}, the least squares error function between the target optimal function \(Q^*(s,a)\) and the current approximation \(Q(s,a)\) is 
\begin{equation}
    \begin{split}
        E(w)&= \frac{1}{2}\left[Q^*(s,a)-Q(s,a)\right]^2 \\
                &= \frac{1}{2}\left[Q^*(s,a)-\sum_i w_i f_i(s,a)\right]^2.
    \end{split}
    \label{eq:lserr}
\end{equation}
%\CY{I removed the sum over $(s,a)$ because that is a much harder problem, and requires optimizing over all $(s,a)$ pairs, which is not what we do. Our RL is a still a greedy algorithm, and optmize at each $(s,a)$ although $w_i$ is global and supposed to be optimal for all $(s,a)$ as we go through different states and take different actions.}
To minimize \eqref{eq:lserr}, we compute the gradient of $E(w)$ formally as follows 
\[\frac{\partial E(w)}{\partial w_i}= - \left[Q^*(s,a)-\sum_j w_j f_j(s,a)\right]\cdot f_i(s,a).\]
The least squares error may be reduced if we move along the negative gradient direction by updating $w_i$'s as
\begin{equation}
w_i \leftarrow w_i + \alpha \cdot \left[Q^*(s,a)-\sum_j w_j f_j(s,a)\right]\cdot f_i(s,a),
\label{eq:wupdate}
\end{equation}
where the learning rate $\alpha$ is simply an appropriately chosen step length.

Note that \eqref{eq:wupdate} can also be written as
\begin{equation}
w_i \leftarrow w_i + \alpha \cdot \left[Q^*(s,a)-Q(s,a)\right]\cdot f_i(s,a)
\label{eq:wupdate2}
\end{equation}
by making use of the definition of $Q(s,a)$ given in \eqref{eq:qsa}.

Because $f_i(s,a)$'s are nonzero for a few $i$'s, only a few $w_i$'s are modified in each step. In fact, if $f_i(s,a)$ is defined by \eqref{eq:fidiff},
only two components of $w_i$'s are update because
\(Q(s,a)\) is defined by
\begin{equation}
	Q(s,a) = w_q - w_p,
	\label{eq:qsaupdatediff}
\end{equation}
where \(a=(p,q)\), \(p \in s\) and $q\notin s$.

However, since we do not know the optimal \(Q^*(s,a)\) in advance, the formulae given 
by \eqref{eq:wupdate} and \eqref{eq:wupdate2} are not computable. To work around this issue, we replace $Q^{\ast}(s,a)$ by a surrogate 
function defined in terms of a local reward and discounted future $Q(s',a')$ values.

A commonly used surrogate for
$Q^{\ast}(s,a)$ is 
\begin{equation}
r + \gamma \cdot \max_{a'} Q(s',a'),
\label{eq:surrogateQ}
\end{equation}
where $r$ is a local reward define in \ref{sec:reward},
and $\gamma$ is an appropriately chosen discount rate. This choice yields the updating formula given in \eqref{eq:updateRule}. 
This choice of the surrogate is likely to be poor in early iterations and episodes, but can gradually converge to the optimal $Q(s,a)$ as the agent visits more states and take more actions.

When $f_i(s,a)$ is defined by \eqref{eq:fisum} and the local reward $r$ is defined by \eqref{eq:rsum}, it is difficult to draw a direct connection between \eqref{eq:wupdate} and 
a gradient descent minimization scheme for obtaining an optimal $Q(s,a)$. However, the updating formula \eqref{eq:wupdate} can still be derived from the updating formula for $Q(s,a)$, i.e.,
\begin{equation}
Q(s,a)\leftarrow Q(s,a)+\alpha \cdot (r+\gamma\max Q(s',a')-Q(s,a)), 
\label{eq:qsaupdate}
\end{equation}
which is widely used in 
$Q$-learning~\cite{sutton2018reinforcement}.

When $f_i(s,a)$ is defined by \eqref{eq:fisum}, $Q(s,a)$ is defined by
\begin{equation}
Q(s,a) = \sum_{i\in s'} w_i-w_p,
\label{eq:qsasum}
\end{equation}
where $s'$ is the state reached after the action $a$ is taken and the action pair is $(p,q)$ which means taking $p$ out of $s$ and adding $q$ into the new $s'$.

Substituting \eqref{eq:qsasum} and 
\eqref{eq:rsum} into \eqref{eq:qsaupdate} yields
\begin{equation}
(\sum_{i\in s'}w_i-w_p)\leftarrow(\sum_{i\in s'}w_i-w_p)+\alpha \cdot \left[ \sum_{i\in s'} r_i+\gamma \max Q(s',a')\sum_{i\in s'} \xi_i^2-(\sum_{i\in s'} w_i-w_p)\right],
\end{equation}
where  $x_1 = (\xi_1, \xi_2, ..., \xi_k)^T$ is the eigenvector corresponding to the smallest eigenvalue $A_1$ associated with the new state $s'$.

Partitioning and regrouping terms by the indices $i$ and $p$, and dividing $\max_{a'}Q(s',a')$ proportionally according to the weighting factor $\xi_i^2$ among all these $k$ terms yields the following updating formula for all $i\in s'$:
\begin{equation}
w_i \leftarrow w_i+\alpha \cdot \left[r_i+\xi_i^2\gamma\max Q(s',a')-w_i\right].
\label{eq:wupdate3}
\end{equation}
For $w_p$, the updating formula is as follows:
\begin{equation}
w_p \leftarrow w_p+\alpha \cdot \left[-w_p\right],
\label{eq:wupdate4}
\end{equation}
which can be regarded as a penalty for the removed row/column.

\subsubsection{Termination criterion}
An RL procedure typically consists of several episodes.  Many steps (actions) are taken in each episode to identify a better state. We terminate each episode either when no action can be found to improve the $Q(s,a)$ function or when a maximum number of steps per episode is reached.

Because RL is a greedy algorithm that makes use of a number of problem dependent heuristics (e.g., the choice of features  and learning and exploration rate.), its convergence is generally not guaranteed.  As a result, a maximum number of episodes is often set to terminate the RL algorithm when the algorithm does not reach convergence after an excessive number of episodes.

When the RL is terminated, we use the weights produced by the training episodes to rank all rows and columns of $A$, and choose $k$ rows and columns associated with $k$ largest $w_i$'s as the final state, which may be different from the state reached at the of the last episode of the RL algorithm.

%% file: results.tex
\section{Numerical examples}
\label{sec:results}
In this section, we demonstrate the effectiveness of the RL algorithm on solving $k$-sparse eigenvalue problems.

\subsection{Test problems}
\begin{figure}[htbp]
    \centering
    \includegraphics[width =0.5\linewidth]{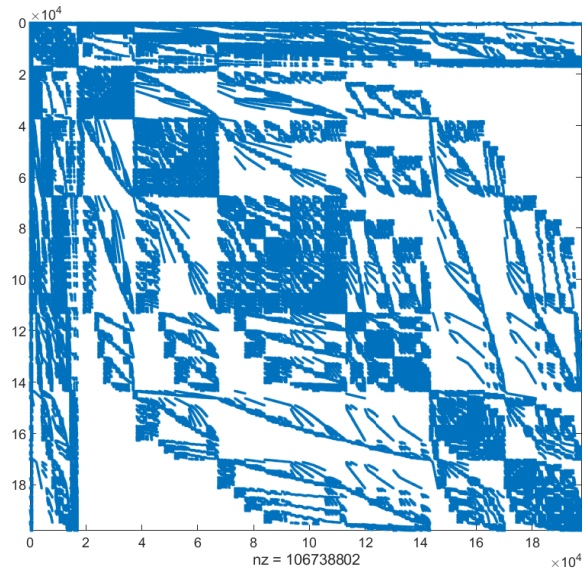}
    \caption{The sparsity of the Li6 Hamiltonian matrix.}
    \label{fig:Li6Nmax6_sparsity}
\end{figure}

We choose two test problems arising from many-body physics applications.
The first problems arises from nuclear structure analysis of the
$^6$Li the isotope, which consists of 3 protons and 3 neutrons. 
The goal is to compute the ground
state of nucleus which is the smallest eigenvalue of the nuclear Schr\"{o}dinger Hamiltonian operator.  The Hamiltonian is approximated in a truncated configuration interaction (CI) space consisting of Slater determinant basis functions that satisfy a truncation cutoff constraint defined in terms of a parameter $N_{\max}$ \cite{caprio20}. We choose 
$N_{\max} = 6$ in our test. 
This choice yields a sparse matrix $A$ of dimension $197,822$. The sparsity pattern of the matrix is shown in Figure~\ref{fig:Li6Nmax6_sparsity}. We label this matrix as Li6Nmax6.

\begin{figure}[htbp]
	\centering
	\includegraphics[width =0.6\linewidth]{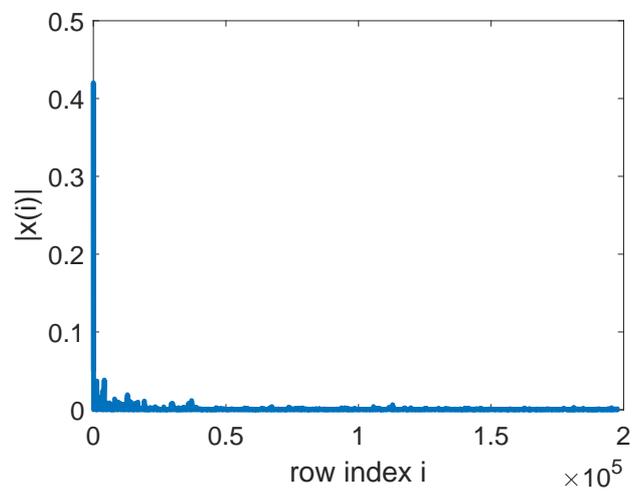}
	\caption{The component magnitude of eigenvector corresponding to the smallest eigenvalue for Li6 Hamiltonian matrix.}
	\label{fig:Li6evec}
\end{figure}

In general, the Slater determinants that satisfy a smaller $N_{\max}$ cutoff tend to contribute more to the ground state, although this is not always true. The true eigenvector associated with smallest eigenvalue of $A$ is shown in Figure~\ref{fig:Li6evec}. We observe that this eigenvector has larger magnitudes in the first few components corresponding to Slater determinants in a small configuration space. But the eigenvector is not strictly localized. Other components of the eigenvector are small but not negligible.

\begin{figure}[htbp]
	\centering
	\includegraphics[width =0.5\linewidth]{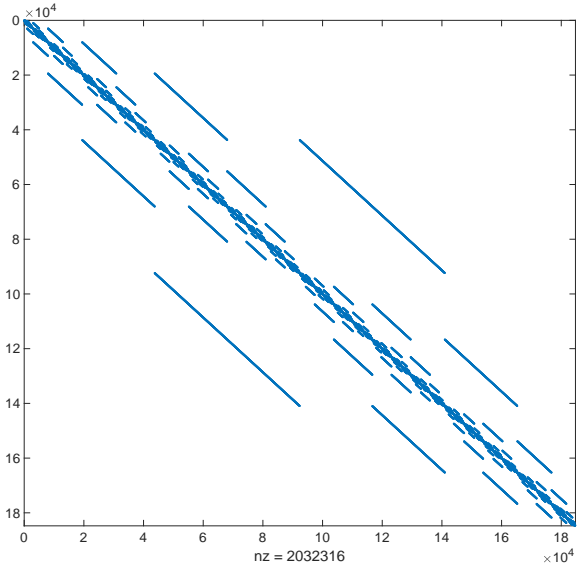}
	\caption{The sparsity of the MBL20 Hamiltonian matrix.}
	\label{fig:MBL20_sparsity}
\end{figure}

\begin{figure}[htbp]
	\centering
	\includegraphics[width =0.6\linewidth]{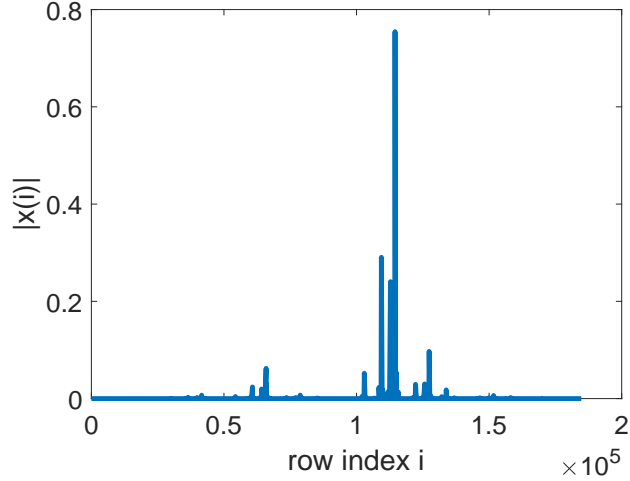}
	\caption{The component magnitude of eigenvector corresponding to the smallest eigenvalue for MBL20 Hamiltonian matrix.}
	\label{fig:MBL20evec}
\end{figure}

The second problem we use for testing originates from the study of many-body localization (MBL)~\cite{luitz2015many,smith2016many,nandkishore2015many}. 
The sparse matrix $A$ represents the Heisenberg spin-1/2 Hamiltonian associated with a disordered quantum spin chain with \(L = 20\) spins and nearest neighbor interactions~\cite{van2020scalable}. 
We name this matrix as MBL20.
The dimension of the MBL20 martrix is $184,756$ and its sparsity structure is shown in Fig~\ref{fig:MBL20_sparsity}. 
This matrix is much sparser, with only $0.006\%$ nonzero elements.
When the strength of the disorder is sufficiently large, 
the eigenvectors of MBL20 exhibits localization features. In this paper, we compute only the smallest eigenvalue and the corresponding eigenvector. Figure~\ref{fig:MBL20evec} shows the magnitudes of different components of this eigenvector.

We set $k$ to 100 for both test problems, which is admittedly small.  Even though the eigenvector of the MBL20 matrix is localized. The number of nonzero components of the eigenvector is actually larger than 1000. 
Nonetheless, by setting $k=100$, we can still demonstrate the features and effectiveness of the RL algorithm.

\subsection{RL options and parameters}
We use two variants of the RL algorithm to solve the $k$-sparse.
In the first variant, which we denote by RL1, the feature vector associated with each state action pair is defined by \eqref{eq:fidiff}. The corresponding $Q(s,a)$ function is defined by \eqref{eq:qsaupdatediff}. The search policy is designed to solve \eqref{eq:policy}. As a result, each action simply swaps out a row in $s$ that has the least $w_i$ with a row outside of $s$ that has the largest $w_j$. Only $w_p$ and $w_q$ are updated by the formula given by \eqref{eq:wupdate}.

In the second variant, which we label by RL2, the feature vector $f(s,a)$ is chosen to be \eqref{eq:fisum}. The corresponding $Q(s,a)$ function is defined by \eqref{eq:qsaupdate}. The alternative search policy given in Algorithm~\ref{alg:policy} is used to select an action. The update of the weighting factors $w_i$'s follows the formula given by \eqref{eq:wupdate3} and \eqref{eq:wupdate4}.

The learning rate for updating $w_i$'s is set to $\alpha=0.5$ for both RL1 and RL2. The discount rate $\gamma$ for RL1 is set to 0.1. For RL2, it is set to $1.0$. RL2 uses an exploration rate that decays exponentially with respect to the number of steps within each episode.

We initialize the state $s$ for both variant to an approximate solution obtained from the greedy algorithm discussed in section~\ref{sec:greedy}.

In RL1, a random starting guess of $s$ is chosen for subsequent episodes, whereas in RL2, the state reached in the previous episode is chosen as the starting state for the next episode.

Table~\ref{tab:algoptions} summarizes the algorithmic and parameter differences between RL1 and RL2.
\begin{table}[H]
	\centering
	\begin{tabular}{|l|c|c|}
		\hline
		            & RL1 & RL2  \\ \hline
		feature representation  &  Eq. \eqref{eq:fidiff}  & Eq. \eqref{eq:fisum}   \\ \hline
		$Q(s,a)$    &  Eq. \eqref{eq:qsaupdatediff} & \eqref{eq:qsaupdate}    \\ \hline
	    policy      &  solve \eqref{eq:policy} & Algorithm~\ref{eq:policy} \\ \hline
		$w_i$ update  &  Eq. \eqref{eq:wupdate} & Eqs. \eqref{eq:wupdate3} and \eqref{eq:wupdate4}  \\ \hline
		learn rate    &  0.5  & 0.5   \\ \hline
		discount rate    & 0.1 & 1 \\ \hline
		exploration rate & N/A &  $e^{-i}$ \\ \hline
	\end{tabular}
	\caption{Algorithmic options and parameters for two different variants of the RL algorithm.}
	\label{tab:algoptions}
\end{table}

\subsection{Convergence}
\begin{figure}[htbp]
    \centering
    \includegraphics[width =0.6\linewidth]{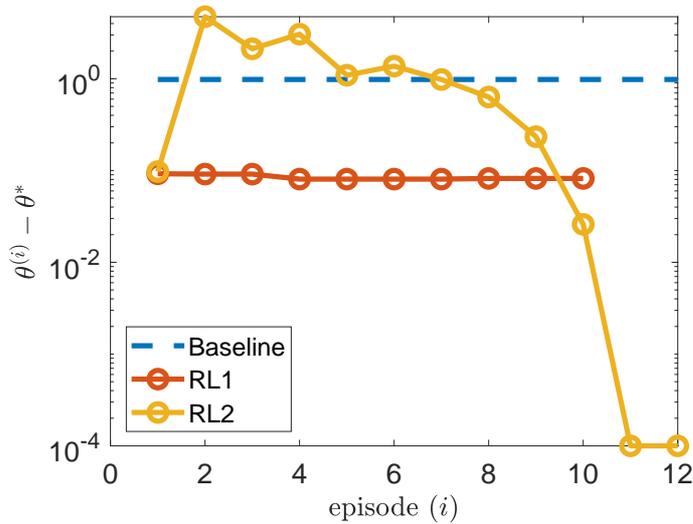}
    \caption{The differences between the best eigenvalue obtained from RL2 and the lowest eigenvalue at the end of each training episode for the \(^6\text{Li}\) matrix from RL1 and RL2.}
    \label{fig:evLi6}
\end{figure}

In Figure~\ref{fig:evLi6}, we show the convergence of the smallest eigenvalue of $A_1$ to the solution of the $k$-sparse eigenvalue problem for the Li6Nmax6 problem in both RL1 and RL2.  We denote the best 
eigenvalue approximation obtained from RL2 by $\theta^*$. Instead of plotting the change of eigenvalues of $A_1$ with respect to the episode number, we plot the difference $\theta^{(i)} - \theta^*+10^{-4}$,  where
$\theta^{(i)}$ is the eigenvalue of $A_1$ obtained at the end of the $i$th episode.
The small constant $10^{-4}$ is added to avoid plotting 0 on a log scale for the last episode of RL2.  The dashed line marks
the difference between the baseline solution $\lambda^b$ and $\theta^*$, where
the baseline solution is obtained by selecting rows and columns of $A$ corresponding to the largest (in magnitude) $k$ components of the eigenvector associated with the smallest eigenvalue of $A$.   
As we indicated in Section~\ref{sec:intro}, any solution that falls below this baseline is considered a ``good" solution.  The solution to the $k$-sparse eigenvalue problem lies below the blue lines. However, we do not know the exact solution to \eqref{eq:evksparse}, which in principle can be obtained by enumerating all possible combinations of $k=100$ rows and columns of $A$ and computing the smallest eigenvalue of the corresponding $A_1$. Therefore, we cannot show how far is $\theta^*$ to the exact solution.

We observe that both RL1 and RL2 can find approximate solutions that are ``good".  
There is very little change in the eigenvalue of $A_1$ in RL1 after the first few episodes. In RL2, the smallest eigenvalue of $A_1$ increases in the first few episodes, but eventually decreases and falls below that produced by RL1. This is likely due to the active space based search policy used in RL2 to explore a wider  range of actions.

We also compared the computed eigenvalues obtained from RL1 and RL2 with a solution obtained by reformulating \eqref{eq:evksparse} as a sparse principal component analysis (PCA) problem and using the GPower$_{l1}$ method presented in \cite{journee2010generalized} to solve this problem. 
In such a reformulation, we try to find
\begin{equation}
x^*=\argmax_{x^Tx\leq 1}\sqrt{x^T(\sigma I-A)^2 x}-\rho\|x\|_{1},
\label{eq:sparsepca}
\end{equation}
where $\sigma$ is a shift chosen to map the smallest eigenvalue of $A$ to the largest eigenvalue of $(\sigma I - A)^2$ and  $\rho$ is  a $L_1$ penalty parameter used to introduce sparsity in an approximate solution to \eqref{eq:sparsepca}\cite{tibshirani1996regression}. The larger the $\rho$, the sparser the approximate solution $x$ will be. The GPower$_{l1}$ method is based on the power iteration. By setting $\rho$ to $0.003$, the GPower$_{l1}$ method produces an approximate solution $x$ that has $100$ elements that are significantly larger than 0 in magnitude.    

Table~\ref{tab:evcomp}
shows that both RL1 and RL2 produce better approximate eigenvalues than that obtained from GPower$_{l1}$.
\begin{table}[H]
	\centering
	\begin{tabular}{|l|c|}
		\hline
		Method         &  Approximate Eigenvalue      \\ \hline 
		RL1            &  -22.9503   \\ \hline
		RL2            &  -23.0441   \\ \hline
		GPower$_{l1}$  &  -22.7243   \\ \hline
		baseline       &  -22.0618  \\ \hline
		Greedy initialization  &  -22.9468 \\ \hline
	\end{tabular}
	\caption{A comparison of approximate eigenvalues obtained from RL1, RL2 and a sparse PCA solver GPower$_{l1}$.}
	\label{tab:evcomp}
\end{table}

\begin{figure}[htbp]
	\centering
	\includegraphics[width =0.6\linewidth]{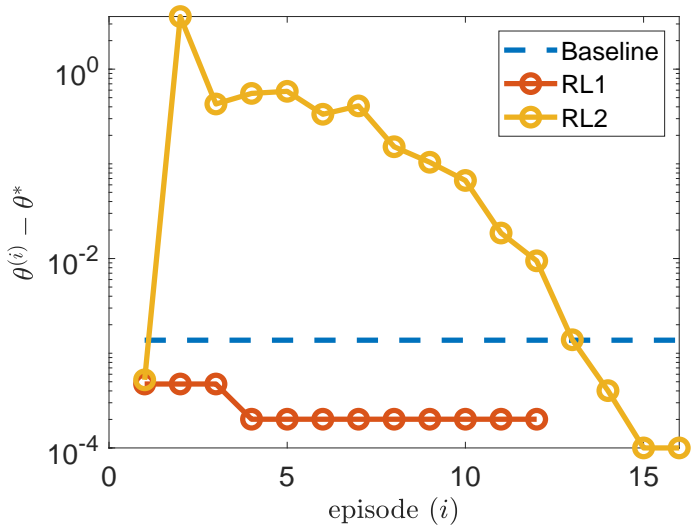}
	\caption{The differences between the best eigenvalue obtained from RL2 and the lowest eigenvalue at the end of each training episode for the MBL20 matrix from RL1 and RL2.}
	\label{fig:evMBL20}
\end{figure}

We observe similar convergence patterns in RL1 and RL2 when they are applied to the MBL20 problem.
Figure~\ref{fig:evMBL20} shows that the smallest eigenvalue of $A_1$ obtained at the end of 
each episode in RL1 changes very little after the third episode. The approximate eigenvalues computed
in RL2 increases above the baseline eigenvalue $\lambda^b$ in the second episode, but then gradually decreases in subsequent
episodes until it reaches the best value reported in Table~\ref{tab:evcomp2}.
Approximate solutions obtained from both RL1 and RL2 are “good”. The difference between
the baseline eigenvalue $\lambda^b$ and $\theta^*$ (reached by RL2) is plotted as a dashed line.
Table~\ref{tab:evcomp2} shows the final approximate eigenvalue from RL1, RL2 and baseline. 
\begin{table}[htbp]
	\centering
	\begin{tabular}{|l|c|}
		\hline
		Method         &  Approximate Eigenvalue      \\ \hline
		RL1            &  -27.9241   \\ \hline
		RL2            &  -27.9242   \\ \hline
		baseline       &  -27.9229   \\ \hline
		Greedy initialization &  -27.9237   \\ \hline
	\end{tabular}
	\caption{A comparison of approximate eigenvalues obtained from RL1 and RL2 for the MBL20 matrix.}
	\label{tab:evcomp2}
\end{table}

\subsection{State evolution}
\begin{figure}[htbp]
	\centering
	\includegraphics[width =0.95\linewidth]{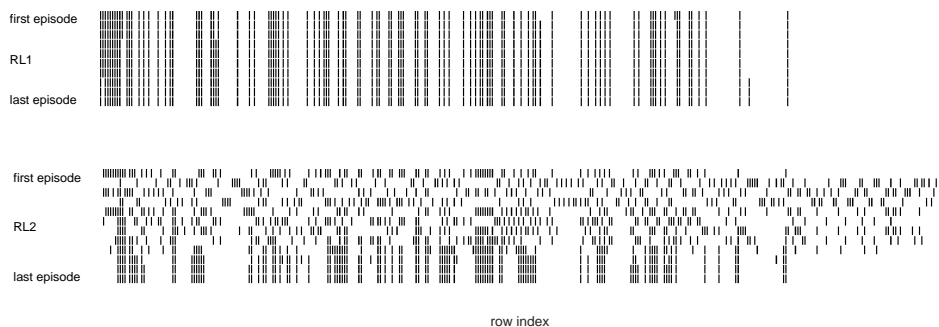}
	\caption{State evolution in RL1 and RL2 for the Li6Nmax6 problem.}
	\label{fig:Li6map}
\end{figure}

In Figures~\ref{fig:Li6map}, we show how the states evolve from one episode to another in RL1 and RL2 respectively. The state $s$ reached at the end of each episode is plotted as shaded rectangles in a horizontal array of rectangles indexed by the row numbers of the matrix $A$. We only plot up to the largest row index that has been selected.

Notice that the state evolution pattern is quite different in RL1 from that in RL2.  The states evolve slowly in RL1. With a good initial guess provided by the greedy algorithm, only a few rows and columns are changed over several episodes of RL1.
The change of states are much more dramatic in RL2. We can see that RL2 explores a much wider range of row/column indices. This is partly because RL2 uses the active space based search policy to choose an action in each step, and partly due to the more frequent use of exploration in the early episodes of the algorithm. Although the states reached in earlier episodes are far from optimal, RL2 eventually converges to a better state than the one identified by RL1, as we have shown in Figure~\ref{fig:evLi6}.

\begin{figure}[htbp]
	\centering
	\includegraphics[width =0.95\linewidth]{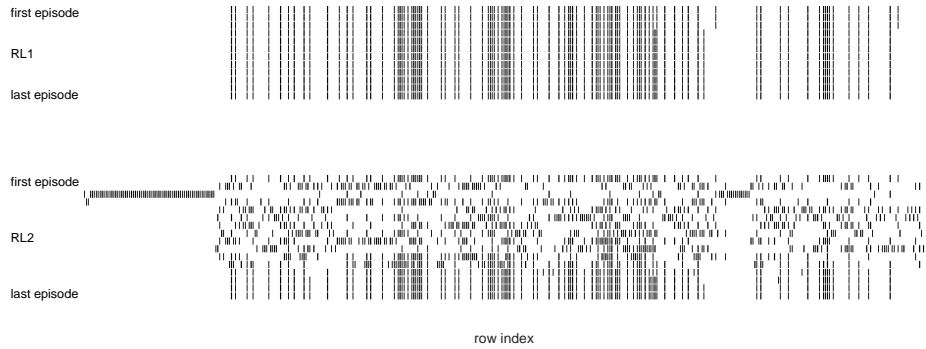}
	\caption{State evolution in RL1 and RL2 for the MBL20 problem.}
	\label{fig:MBL20map}
\end{figure}

For MBL20, as shown in Figures~\ref{fig:MBL20map}, RL1 again converges in a few episodes starting from an initial state produced from the greedy algorithm.  The states reached in the first several episodes changes quite a bit again in RL2 reflecting the use of active space based policy and more extensive exploration. But eventually the algorithm settles down to a subset of rows and columns that corresponds to the localized region of the eigenvector. Due to the more localized nature of the eigenvector for this problem, the difference between the states reached by RL1 and RL2 is relatively small. This is consistent with the approximate eigenvalues reported in Table~\ref{tab:evcomp2}.

\subsection{Weights at Convergence}
\begin{figure}[H]
    \centering
    \includegraphics[width =0.6\linewidth]{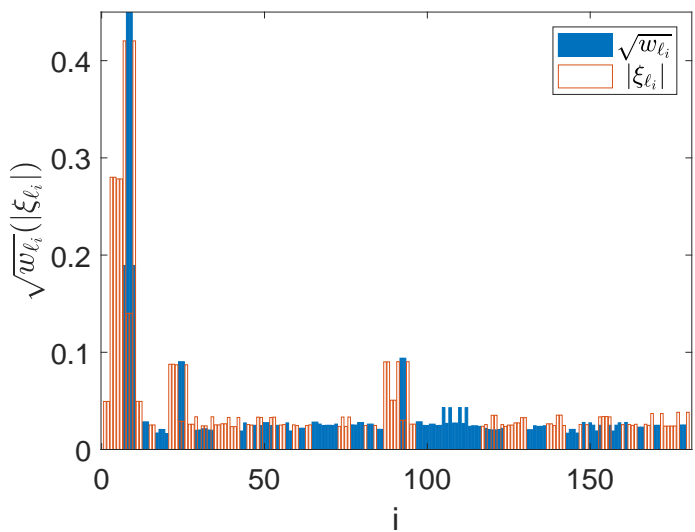}
    \caption{The absolute value of the lowest eigenvector \(x\) components of \(^6\text{Li}\) overlayed with  of weights \(w\) from RL1 of the corresponding rows. The weight $w$ are rescaled to a unit vector for compare.}
    \label{fig:Li6weights}
\end{figure}

\begin{figure}[H]
    \centering
    \includegraphics[width =0.6\linewidth]{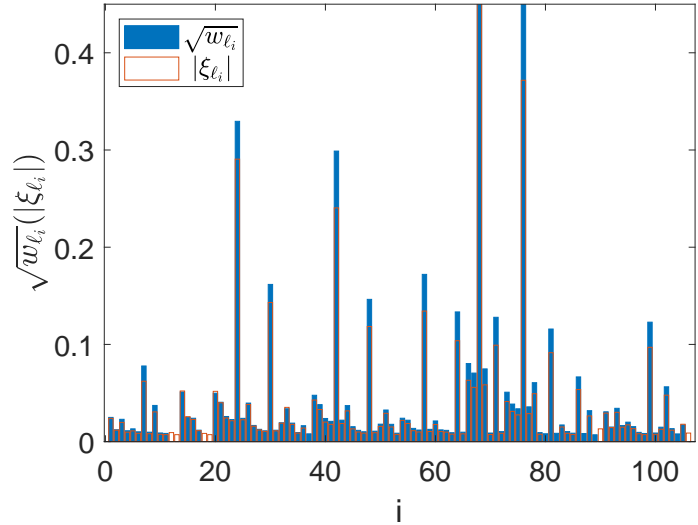}
    \caption{The absolute value of the lowest eigenvector \(x\) components of MBL overlayed with the square root of weights \(w\) from RL2 of the corresponding rows. The weight $w$ are rescaled to a unit vector for compare.}
    \label{fig:MBLweights}
\end{figure}

In Figures~\ref{fig:Li6weights} and~\ref{fig:MBLweights}, we plot 
$\sqrt{w_i}$'s for the largest $k=100$ weights produced when RL2 is terminated.  We compare the location of these weights with the locations of the largest $k=100$ components (in magnitude) of the desired eigenvector of $A$.
To make it easier to see the difference between the row indices selected by RL2 and the row indices corresponding to the largest eigenvector components (baseline solution),
we collect the indices $\mathcal{I}=\{i\}$ associated with the largest $w_i$'s and the indices $\mathcal{J}=\{j\}$ associated with the largest eigenvector components.
We then take the union of $\mathcal{I}$ and $\mathcal{J}$, $\mathcal{K} = \mathcal{I}\cup \mathcal{J}$, and sort the indices in $\mathcal{K}$ in an increasing order and denote them by
\[
\ell_1 < \ell_2 ... < \ell_{|\mathcal{K}|},
\]
where $|\mathcal{K}|$ is the size of $\mathcal{K}$.
We plot $\sqrt{w_{\ell_i}}$ as a solid blue bar over $i$ if $\ell_i \in \mathcal{I}$, and an eigenvector component $|\xi_{w_{\ell_i}}|$ as a red empty bar over $i$ if $\ell_i \in \mathcal{J}$.

We observe from Figure~\ref{fig:Li6weights} that
only a small subset of rows selected by RL2 for the Li6Nmax6 problem overlap with the largest 100 elements of the desired eigenvector of $A$. Because the solution obtained by RL2 corresponds to a much smaller eigenvalue, it is a much better solution than the baseline solution.

For this problem, which is not strictly localized, RL appears to be very effective in finding a very good solution even though we can not verify that this solution is the solution of \eqref{eq:evksparse}.

Figure~\ref{fig:MBLweights} shows that, for MBL20, there is a significant overlap between the rows of $A$ selected by RL2 and those that correspond to 100 largest eigenvector components associated with the smallest eigenvalue of $A$. To some extent, this is not surprising because the desired eigenvector of the MBL20 problem has a clear localization feature shown in Figure~\ref{fig:MBL20evec}.

%% file: conclusion.tex
\section{Conclusion}
We showed how a $k$-sparse eigenvalue problem can be solved via a RL algorithm.
We described how to represent states ($s$), actions ($a$), local rewards, and the global expected 
return (also known as the $Q(s,a)$ function), which are the basic ingredients of a RL algorithms, 
for the $k$-sparse eigenvalue problem.  Each state simply consists of indices of the $k$ rows and columns
selected from $A$. Each action involves removing one index from $s$ and adding another index outside of $s$
to form a new state $s'$.  The most challenging problem for devising an effective and 
efficient RL algorithm is to construct an appropriate representation of the $Q(s,a)$ function, 
which values the suitability of taking the action $a$ at the state $s$, so that it can be 
updated efficiently during a learning process, especially when both the number of states and actions 
that we can take in using RL to solve the $k$-sparse eigenvalue problem is extremely large.  
In this paper, we choose to represent $Q(s,a)$ as a linear
combination of a number of feature components defined in terms of row and column indices in the 
current or the next states. This linear representation is sufficiently simple so that the update
of $Q(s,a)$ can be realized by modifying the weighting factors in the linear combination.
We presented two strategies (policies) for choosing the next action based on the information provided
in $Q(s,a)$ and component-wise perturbation analysis. In particular, we proposed an active space 
based policy that search within a subset of candidate actions that can lead to a significant
reduction of the smallest eigenvalue of the selected submatrix. 
We tested the RL algorithm on two examples originating from many-body physics. One of the 
problems has nice localization properties whereas the other is not strictly localized.
We demonstrated the effectiveness of RL on both problems.
Although these problems are still relatively small, and we chose a relative small $k$ so that
the RL algorithms can terminate in a handful of episodes, they are good benchmarks for testing
the robustness and effectiveness of the RL algorithm.  Clearly, more tests involving larger
problems and larger $k$'s need to be performed in the future to develop strategies for setting a 
number of RL parameters such as learning rate, exploration rate and discount rate. 
As the problem size becomes larger, many other computational considerations such as 
fast local reward computation, the parallelization of active space search and exploration need to 
be developed. 
Furthermore, we may also consider using a nonlinear representation of the $Q(s,a)$ function 
perhaps via a deep neural network to further improve the predictability of $Q(s,a)$ and consequently
the search policy and expand the space of action by including actions that involve removing and 
adding multiple rows and columns of $A$ from and to $A_1$.
In addition to comparing with baseline solution, we will also compare RL solution to solutions obtained 
from Monte Carlo sampling based methods.